\documentclass{optica-article}

\journal{opticajournal} % use for journal or Optica Open submissions

\articletype{Research Article}

\usepackage{lineno}
\usepackage[utf8]{inputenc}

\usepackage{graphicx} % Required for inserting images
\usepackage{amsmath, amsfonts}
\usepackage{physics}
\usepackage{bm}
\usepackage{xcolor}

\newcommand{\vE}{\vb{E}}
\newcommand{\vP}{\vb{P}}

\renewcommand{\vr}{\vb{r}}
\newcommand{\vx}{\vb{x}}

\newcommand{\G}{\mathbb{G}}

\newcommand\numthis{\stepcounter{equation}\tag{\theequation}}

\renewcommand{\eqref}[1]{Eq.~(\ref{#1})}
\newcommand{\figref}[1]{Fig.~\ref{#1}}

\begin{document}

\title{Sum-of-Squares Bounds on Surface-Enhanced Raman Scattering}

\author{Pengning Chao, \authormark{1,*} Ian M. Hammond,\authormark{2} and Steven G. Johnson\authormark{1}}

\address{\authormark{1} Department of Mathematics, Massachusetts Institute of Technology, Cambridge, USA \\
\authormark{2} Department of Electrical Engineering and Computer Science, Massachusetts Institute of Technology, Cambridge, USA}

\email{\authormark{*} pchao827@mit.edu}

\begin{abstract}
   Surface-enhanced Raman scattering (SERS) is a critical tool for chemical sensing and spectroscopy, and a key question is how to optimally design nanostructures for maximizing SERS. 
   We present fundamental limits on spatially-averaged SERS via periodic metasurfaces, derived using sum-of-squares (SOS) programming. This work represents the first use of SOS techniques to optics, overcoming difficulties that prior bounding techniques have with regards to non-linear photonic processes with higher order figures of merit. Our bounds on the $\int \lVert \mathbf{E} \rVert^4 \dd \vr$ SERS enhancement factor for 2D examples demonstrate remarkable tightness when compared with inverse-designed dielectric and metallic structures for both electrical field out-of-plane ($E_z$) and in-plane ($H_z$) polarizations. We show that delocalized high-Q guided modes can achieve significant, theoretically diverging SERS enhancement even in the presence of material loss. For metallic structures, we demonstrate a fundamental performance limitation for $E_z$ polarized drive fields due to surface plasmon excitation restrictions. By varying the separation between Raman-active molecules and the metasurface design region, we also find material-dependent bounds on the maximum strength of field singularities. Our results offer insights into optimal metasurface design strategies for enhancing light-matter interactions, and our methodology may be adapted to the study of other nonlinear photonics design problems. 
\end{abstract}

\section{Introduction}

We present the first ever bounds on a quartic electromagnetic figure of merit (FOM) over all possible geometries, specifically the $\int \norm{\vE}^4 \dd\vr$ FOM for surface-enhanced Raman scattering (SERS) via photonic metasurfaces~\cite{leruRigorousJustification2006,yaoDesigningStructures2023}. The bounds represent the first application of sum-of-squares (SOS) programming~\cite{parriloStructuredSemidefinite2000,lasserreGlobalOptimization2001,blekhermanSemidefiniteOptimization2013} in optics and provide a general approach for extending recent photonic design bounding techniques~\cite{chaoPhysicalLimits2022,angerisHeuristicMethods2021,gertlerManyPhotonic2025} from linear to nonlinear optics. 
We compute the bounds for 2D examples and compare them with the performance of inverse-designed metasurfaces~\cite{yaoDesigningStructures2023} made out of dielectrics or metals, for both $E_z$ and $H_z$ polarizations. We find that our bounds are remarkably tight, often coming within a small constant factor of the inverse designs. Both bounds and inverse designs identify a theoretical opportunity for large SERS enhancement---even in the presence of substantial material loss---through the use of delocalized \emph{arbitrarily} high-$Q$ guided modes~\cite{strekhaSuppressingElectromagnetic2024}. Furthermore, the bounds prove the importance of incident light polarization for SERS via metallic structures, showing a low performance ceiling for the~$E_z$ polarization, related to boundary condition requirements for exciting surface plasmon polaritons (SPPs)~\cite{maierPlasmonicsFundamentals2007a} in metals. By varying the separation between Raman active molecules and the metasurface design region, we also obtained numerical data on the strongest field singularity~\cite{felsenRadiationScattering2003,idemenConfluentTip2003,budaevElectromagneticField2007,kottmannSpectralResponse2000} theoretically achievable via nanostructuring. 

Raman scattering---the inelastic scattering of photons---is a principle method for probing the energy structure of molecules, with important applications in chemical sensing and spectroscopy~\cite{garrellSurfaceenhancedRaman1989,kneippSurfaceEnhancedRaman2006,jonesRamanTechniques2019}. SERS boosts the Raman signal by placing the sample in the vicinity of nanostructures that strongly increase the local photonic density of states (both concentrating the incident wave and enhancing the emission process); this is commonly implemented with metallic nanoparticles supporting plasmonic resonances~\cite{moskovitsSurfaceenhancedSpectroscopy1985,langerPresentFuture2020}. Recently, photonics inverse design~\cite{moleskyInverseDesign2018} has been used to develop nanostructures that maximize SERS from both fixed hotspot locations~\cite{christiansenInverseDesign2020,panTopologyOptimization2021} as well as from dispersed molecules in a fluid~\cite{yaoDesigningStructures2023}. In this manuscript, we are particularly interested in the latter ``distributed'' emission process due to both its theoretical challenge and practical relevance for chemical detection. 

\begin{figure}[ht!]
    \centering
    \includegraphics[width=0.5\textwidth]{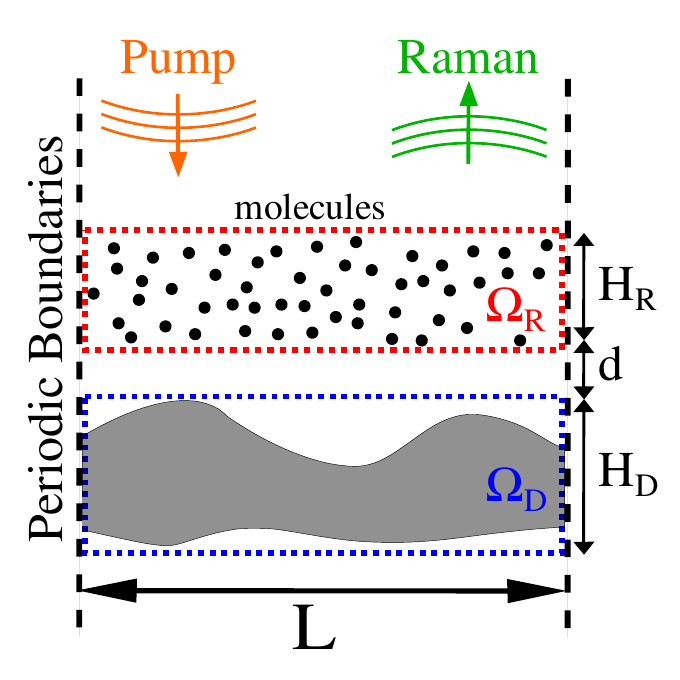}
    \caption{Schematic of the spatially-averaged SERS setting investigated in this work: Raman scattering of a normal incidence pump planewave is enhanced in the normal emission direction through the design of a periodic metasurface. \label{fig:suspended_schematic}}
\end{figure}

Our distributed-emitter Raman scenario is illustrated in \figref{fig:suspended_schematic}: a periodic metasurface with unit-cell period~$L$ is engineered (optimizing the geometric arrangement of a given dielectric or metallic material) within the design region $\Omega_D$ to maximize the Raman emission from molecules freely distributed within the sample region $\Omega_R$. The maximized inverse-design objective is the quartic $\int_{\Omega_R} \norm{\vE}^4 \dd\vr$, where $\vE$ is the electric field from a normally incident pump planewave, a standard SERS FOM that can be derived using reciprocity~\cite{leruRigorousJustification2006,yaoDesigningStructures2023}; mathematically, this is proportional to the average Raman emission in the normal direction from isotropic Raman molecules distributed uniformly over $\Omega_R$. In order to obtain finite upper bounds, we consider a situation in which the Raman molecules ($\Omega_R$) are separated by a distance $d > 0$ from the metasurface design region ($\Omega_D$), as otherwise an arbitrarily large response can theoretically be attained by structuring the material into a specifically-angled sharp tip that supports a field singularity at its apex~\cite{andersenFieldBehavior1978,idemenConfluentTip2003,budaevElectromagneticField2007}; these singularities are discussed further in Sec.~\ref{sec:resultsVaryd}. Such singularities are regularized physically by the presence of nonlocality in the response to material polarization at the nanometer scale~\cite{mcmahonNonlocalOptical2009,ciraciProbingUltimate2012,gubbinOpticalNonlocality2020}, and sharp-tip structures can be avoided at the design level by imposing minimum lengthscale constraints on the geometry~\cite{zhouMinimumLength2015a,vercruysseAnalyticalLevel2019a,hammondPhotonicTopology2021,schubertInverseDesign2022}. While directly incorporating nonlocal polarizability or fabrication lengthscale constraints into bound formulations is still an open problem~\cite{chaoPhysicalLimits2022}, the $d>0$ setting allows us to sidestep these issues and focus our attention on large-area enhancement mechanisms as opposed to unphysical sharp-tip singularities. 
The positive separation is also useful from a computational perspective because it makes the Green's function connecting $\Omega_D$ to $\Omega_R$ approximately low rank~\cite{hackbuschHierarchicalMatrices2015,polimeridisFluctuatingVolumecurrent2015}, as discussed in more detail below.

Although inverse design is an effective tool for achieving high performance, the resulting structures are often complex and non-intuitive, with practical difficulties in determining how close the designs are to global optimality~\cite{moleskyInverseDesign2018}. This has motivated recent major developments in the study of photonic fundamental limits: there is currently a general framework for computing bounds to linear photonics problems where the design objective is a quadratic function of the fields~\cite{angerisHeuristicMethods2021,chaoMaximumElectromagnetic2022,gertlerManyPhotonic2025}. This framework has been applied successfully to many important problems, such as scattering/absorption cross sections~\cite{gustafssonUpperBounds2020,moleskyGlobal$mathbbT$2020,kuangComputationalBounds2020}, the local density of states~\cite{shimFundamentalLimits2019,chaoMaximumElectromagnetic2022}, waveform shaping~\cite{molesky$mathbbT$OperatorLimits2021,shimFundamentalLimits2021}, and optical forces~\cite{venkataramFundamentalLimits2020,strekhaTraceExpressions2024}. In contrast, bounds for nonlinear photonics design problems such as SERS are much less explored, with the few existing results on second-harmonic generation~\cite{mohajanFundamentalLimits2023} and single-molecule SERS~\cite{michonLimitsSurfaceenhanced2019,amaoloPhysicalLimits2024} based upon factoring non-linear photonic FOMs into products of linear photonics FOMs, which may result in excessively loose bounds if the coupling between the linear design problems is not properly accounted for~\cite{amaoloPhysicalLimits2024}. Moreover, for the distributed-SERS quartic FOM such a factorization is not even possible. We overcome this challenge using SOS programming, a powerful technique in polynomial optimization capable of handling general objective functions that are beyond quadratic in order~\cite{parriloStructuredSemidefinite2000,lasserreGlobalOptimization2001,blekhermanSemidefiniteOptimization2013}. To mitigate the increased computational cost associated with SOS programming, we use a low-rank approximation of the vacuum Green's function~\cite{hackbuschHierarchicalMatrices2015,polimeridisFluctuatingVolumecurrent2015} to reduce the degrees of freedom needed for expressing the fields. 

While this paper focuses specifically on bounding SERS, the generality of the SOS programming approach should allow it to be adapted for studying other nonlinear photonics FOMs that are not necessarily factorizable. Given growing interest in nonlinear photonics inverse design~\cite{linCavityenhancedSecondharmonic2016,sitawarinInversedesignedPhotonic2018a,hughesAdjointMethod2018,yangInversedesignedSilicon2023,mannInverseDesign2023a,stichInverseDesign2024}, we anticipate that bounds based on SOS programming will offer crucial guidance on how to optimally engineer photonic devices for enhancing nonlinear light-matter interactions.  

\section{Problem Setting}
We consider the problem of increasing the spatially averaged Raman scattering signal from target molecules dispersed in a fluid through the design of a periodic metasurface, as depicted schematically in \figref{fig:suspended_schematic}. For simplicity, we assume that the molecule's Raman polarizability tensor $\bm{\alpha}$~\cite{kneippSurfaceEnhancedRaman2006} is isotropic and that the molecules are uniformly randomly distributed in a region $\Omega_R$ of space. The molecules are pumped by a normally incident planewave, and we seek to maximize the collective Raman emission back into the normal direction. For simplicity, we assume reciprocal materials and a negligible frequency shift between pump and Raman emission, in which case the relevant FOM is $\int \norm{\vb{E}(\vr)}^4\dd\vr$, where $\vE$ is the total electric field~\cite{leruRigorousJustification2006,yaoDesigningStructures2023}. Designing a SERS metasurface can thus be formulated as the following structural optimization problem:
\begin{subequations}
\begin{equation}
    \max_{\epsilon(\vr)} \quad \int_{\Omega_R} \norm{\vE(\vr)}^4 \,\dd\vr \, ,
    \label{eq:structOptObj}
\end{equation}
given the constraints
\begin{align}
    &\curl \curl \vb{E}(\vb{r}) - \omega^2 \epsilon(\vb{r}) \vb{E}(\vb{r}) = i\omega\vb{J}(\vb{r}), \label{eq:structOptMaxwell}\\
    &\epsilon(\vb{r}) = 
    \begin{cases}
    1 \text{ or } 1+\chi & \vb{r} \in \Omega_D \\
    1 &  \vb{r} \notin \Omega_D
    \end{cases} \, ,
\end{align}
\label{eq:structOpt}
\end{subequations}
where we have used dimensionless units $\epsilon_0=\mu_0=1$, $\omega$ is both the pump and Raman emission frequency, $\chi$ is the susceptibility of the metasurface material and $\epsilon(\vr)$ is the permittivity distribution. 
The metasurface is restricted to lie within a design region $\Omega_D$ while the figure of merit is evaluated over a Raman scattering region $\Omega_R$, with the two regions separated by a distance $d > 0$; this finite separation is introduced to prevent field singularities at sharp tips causing a divergence in the FOM~\cite{hammondDesigningStructures2024}. \eqref{eq:structOpt} can be approached using inverse design techniques such as topology optimization (TO)~\cite{yaoDesigningStructures2023}; however, because \eqref{eq:structOpt} is both infinite dimensional and non-convex, inverse design generally finds local optima with no guarantee of global optimality~\cite{moleskyInverseDesign2018}, and thus only yields a \emph{lower} bound on the attainable performance. An obvious and important question is how to obtain a complementary \emph{upper} bound, in order to determine the potential for future improvement.

\section{SOS Formulation of Distributed Raman Bounds}
A general approach to deriving such fundamental limits~\cite{chaoPhysicalLimits2022} is to formulate a related optimization problem in terms of the induced polarization field within the structure $\vP$, where the geometric details of $\epsilon(\vr)$ are contained implicitly in $\vP$ and the fields can be expressed using a scattering-theory framework involving the vacuum dyadic Green's function $\G(\vr,\vr')$, satisfying
\begin{equation}
    \curl\curl\G(\vr,\vr') - \omega^2 \G(\vr,\vr') = \omega^2 \hat{\vb{I}} \delta(\vr-\vr') \, 
\end{equation}
where $\hat{\vb{I}}$ is the unit dyad. Specifically for bounding the structural optimization problem (\ref{eq:structOpt}), we consider a relaxation (subset) of the full Maxwell constraints:
\begin{subequations}
    \begin{align}
        &\max_{\vP} \quad f(\vP) \equiv \norm{\vE_i + \G_{RD} \vP}_{\Omega_R,4}^4 \label{eq:fieldObj}\\
        & \text{s.t.} \quad C(\vP) \equiv \vE_i^\dagger \vP - \vP^\dagger (\chi^{-1\dagger} - \G_{DD}^\dagger) \vP = 0 \, ,\label{eq:fieldCplxCstrt} 
    \end{align}
    \label{eq:fieldOpt}
\end{subequations}
where $\G_{RD}$ and $\G_{DD}$ are operator sub-blocks of $\G$ connecting sources in $\Omega_D$ to fields in $\Omega_R$ and $\Omega_D$ respectively, and $\dagger$ represents conjugate-transposition of vector fields and operators over $\Omega_D$. 
The SERS FOM is now expressed as the fourth power of the $L_4$ norm over $\Omega_R$ of the total field $\vE$ given as the sum of the incident planewave $\vE_i$ and the scattered field $\G_{rd} \vP$. The constraint $C(\vP)$ is a quadratic function of $\vP$ and expresses power conservation: in particular, if we split $C(\vP) = C_R(\vP) + i C_I(\vP)$, the imaginary part $C_I(\vP)=0$ expresses conservation of resistive power and the real part $C_R(\vP)=0$ expresses conservation of reactive power~\cite{moleskyGlobal$mathbbT$2020,chaoPhysicalLimits2022}. Given that power conservation is just part of the physics prescribed by Maxwell's equations (\ref{eq:structOptMaxwell}), the field optimization problem (\ref{eq:fieldOpt}) is less restrictive than \eqref{eq:structOpt} so at $\overline{\vP}$, the global optimum of \eqref{eq:fieldOpt}, the FOM $f(\overline{\vP})$ is an upper bound to \eqref{eq:structOpt}. Prior work on fundamental limits in photonics has predominantly focused on the case where the objective is also a quadratic function of $\vP$, since this results in a quadratically constrained quadratic program~\cite{parkGeneralHeuristics2017a} which can be further bounded using techniques such as Lagrangian duality~\cite{chaoMaximumElectromagnetic2022} or equivalently semi-definite lifting~\cite{parkGeneralHeuristics2017a,angerisHeuristicMethods2021,gertlerManyPhotonic2025}. Unfortunately, our given objective is \textit{quartic}, and these techniques either do not apply or yield an infinite upper bound, as discussed below. 

\subsection{The generalized Lagrangian dual}
To obtain non-trivial bounds, we formulate a generalized version of the Lagrangian dual problem, expressed in epigraph form~\cite{boydConvexOptimization2004,blekhermanSemidefiniteOptimization2013} as
\begin{subequations}
    \begin{align}
        & \min_{\substack{t \in \mathbb{R} \\ m(\vP) \text{ is quadratic} } } \quad t \\
        \text{s.t.} \quad & t - \underbrace{( f(\vP) + m(\vP) C(\vP) )}_{\text{generalized Lagrangian}} \text{ is non-negative for all }\vP.  \label{eq:nonnegCstrt}
    \end{align}
    \label{eq:generalizedLag}
\end{subequations}
Note that the generalized dual (\ref{eq:generalizedLag}) bounds $f(\overline{\vP})$ for any choice of $m(\vP)$, since $C(\overline{\vP})=0$ and \eqref{eq:nonnegCstrt} together imply that the dummy variable $t \geq f(\overline{\vP})$; this property is referred to as weak duality~\cite{boydConvexOptimization2004}. The standard Lagrangian dual is obtained if the multiplier $m$ is restricted to be a scalar; differing from the standard dual, here we have promoted $m$ to a quadratic polynomial $m(\vP)$.  This promotion is necessary since the quartic growth of $f(\vP)$ for large $\norm{\vP}$ will dominate over the quadratic $m C(\vP)$, resulting in a trivial bound of $+\infty$ for all scalar $m$. In contrast, $m(\vP) C(\vP)$ is itself quartic, and judicious choice of $m(\vP)$ yields a finite bound. 

The non-negativity constraint (\ref{eq:nonnegCstrt}) is difficult to handle computationally: determining whether a general multivariate polynomial is non-negative is an NP-hard problem for polynomial degree $\geq 4$~\cite{ahmadiNPhardnessDeciding2013,blekhermanSemidefiniteOptimization2013}. This is resolved by replacing non-negativity with a more restrictive sufficient condition (hence an upper bound) that the polynomial is a sum-of-squares (SOS) of lower-degree polynomials:
\begin{subequations}
    \begin{align}
        & \min_{\substack{t \in \mathbb{R} \\ m(\vP) \text{ is quadratic} } } \quad t \\
        \text{s.t.} \quad & t - ( f(\vP) + m(\vP) C(\vP) ) \text{ is SOS}. \label{eq:SOSCstrt}
    \end{align}
    \label{eq:SOSgeneralizedLag}
\end{subequations}
Due to the special structure of SOS polynomials, \eqref{eq:SOSgeneralizedLag} can be mapped to an equivalent semi-definite program (SDP) and solved accordingly~\cite{blekhermanSemidefiniteOptimization2013}. While the SOS condition seems too restrictive at first glance, in practice it often produces surprisingly tight convex relaxations of general polynomial optimization problems~\cite{blekhermanSemidefiniteOptimization2013,parriloMinimizingPolynomial2001}, and as we shall see in the Results section it works well for the Raman problem.

\subsection{Efficient numerics exploiting low rank of $\G_{RD}$}
\begin{figure}[htp]
\centering
    \includegraphics[width=0.75\linewidth]{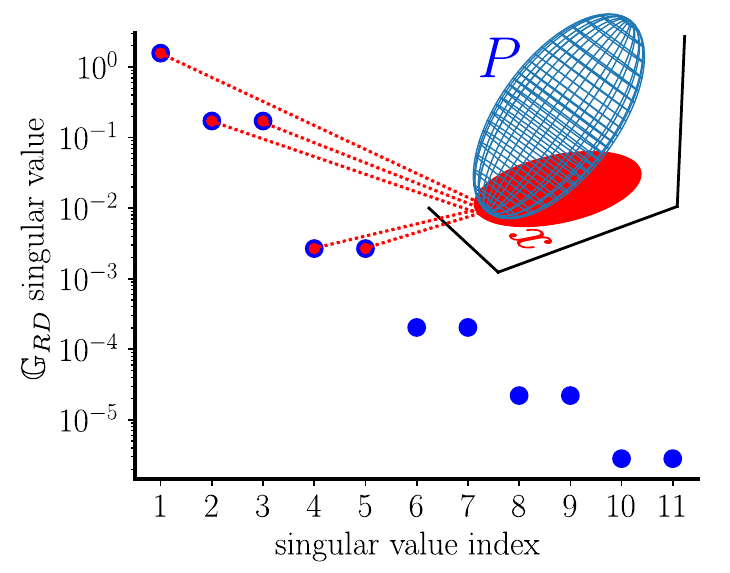}
    \caption{\label{fig:Grdsv}  Plot of the singular values of $G_{RD}$ with $H_D = H_R = 0.5\lambda$ and $d=0.2\lambda$. Given finite separation between $\Omega_R$ and $\Omega_D$, $G_{RD}$ is effectively low-rank with the singular values decaying exponentially. To reduce the computational cost, we restrict SOS analysis to the subspace spanned by the dominant singular vectors, marked schematically in red. Inset: equality constraints (blue ellipsoid surface) for $P$ are projected down into the dominant subspace as inequality constraints (red filled ellipsoid). }
\end{figure}

The final hurdle to getting numerical results is the size of the SOS program (\ref{eq:SOSgeneralizedLag}): if the discretized $\vP$ vector is of length $N$, the quadratic multiplier $m(p)$ has $N^2$ free coefficients. A local discretization such as finite-difference frequency domain (FDFD) or finite elements (FEM) will have $N\gtrsim 10^3, N^2 \gtrsim 10^6$ for 2D wavelength scale domains, making \eqref{eq:SOSgeneralizedLag} intractable for standard SDP-based SOS solvers~\cite{ahmadiDSOSSDSOS2019}. We instead exploit the low-rank property of $\G_{RD}$ when the separation $d$ is nonzero~\cite{hackbuschHierarchicalMatrices2015,polimeridisFluctuatingVolumecurrent2015} and take a singular value decomposition $\G_{RD} = \sum_{j=1}^\infty s_j \vb{u}_j \vb{v}_j^\dagger$ where the singular values $\{s_j\}$ are given in non-increasing order, and the left and right singular vectors $\{\vb{u}_j\}$ and $\{\vb{v}_j\}$ form orthonormal bases for vectors fields over $\Omega_R$ and $\Omega_D$, respectively. See \figref{fig:Grdsv} for a plot of $\{s_j\}$ given a specific design and Raman region specification; in this plot and the results in the next section we will specify lengths in units of the vacuum wavelength $\lambda = 2\pi c/ \omega$. Given the rapid decay of $s_j$ as $j$ increases, $\{\vb{v}_j\}$ gives an efficient basis for representing $\vP$ where we can focus only on the subspace restricted to $j \leq q$ where $q$ is a pre-selected index cutoff. The triangle inequality for the $L_4$ norm on $f(\vP)$ gives
\begin{align*}
    f(\vP) &= \norm{\vE_i + \sum_{j=1}^q s_j (\vb{v}_j^\dagger \vP) \vb{u}_j + \sum_{j=q+1}^\infty s_j (\vb{v}_j^\dagger \vP) \vb{u}_j}_4^4 \\
    &\leq \left( \norm{\vE_i + \sum_{j=1}^q s_j (\vb{v}_j^\dagger \vP) \vb{u}_j}_4 + \norm{\sum_{j=q+1}^\infty s_j (\vb{v}_j^\dagger \vP) \vb{u}_j}_4  \right)^4. \numthis \label{eq:L4triangle}
\end{align*}
The first term in \eqref{eq:L4triangle} is the dominant contribution, which we shall bound using SOS programming as detailed above. For sufficiently large $q$, the contribution from the second term is small, and can in practice be neglected once bounds on the first term have converged with respect to $q$. For the 2D results presented, the bounds generally saturate with $q < 20$, and the resulting SOS programs can be solved in a few minutes on a single multi-core PC (SI). 

Let $\vx$ be the coefficients of $\vP$ in the dominant subspace with $x_j = \vb{v}_j^\dagger \vP$. We shall abuse notation somewhat and replace $\vP$ with $\vx$ as the argument for objective and constraints: in particular, $f(\vx)$ is the fourth power of the first term in \eqref{eq:L4triangle} and a quartic polynomial. Equality constraints $C(\vP)=0$ with a positive definite quadratic form become convex quadratic inequality constraints $C(\vx) \geq 0$ when restricted to the $\vx$ subspace; geometrically this corresponds to projecting a high dimensional ellipsoid surface down into a lower dimension subspace, resulting in a filled-in ellipsoid (SI). $C_I(\vP)$ is such a constraint, with its quadratic operator being the positive definite $\Re{-i (\chi^{-1\dagger} - \G_{dd}^\dagger)}$~\cite{moleskyGlobal$mathbbT$2020}, where $\Re\{\mathbb{A}\} = \frac{\mathbb{A}+\mathbb{A}^\dagger}{2}$ is the real (Hermitian) part of an operator $\mathbb{A}$. 
$C_R(\vP)$ is not such a constraint but we can still incorporate it by forming linear combinations $C_\gamma(\vP) = C_I(\vP) + \gamma C_R(\vP)$ that are. We can thus evaluate bounds on \eqref{eq:structOpt} incorporating conservation of both resistive and reactive power through the following optimization:
\begin{subequations}
\begin{align}
    &\min_{\gamma \in \mathbb{R}} \; \min_{\substack{t \in \mathbb{R} \\ m_I(\vx), m_\gamma(\vx) \text{ are quadratic} } } \quad t \numthis \label{eq:finalBoundForm} \\
    \text{s.t.} \quad  &t - (f(\vx) + m_I(\vx) C_I(\vx) + m_\gamma(\vx) C_{\gamma}(\vx)) \quad \text{is SOS}, \\
    & m_I(x) \quad \text{is SOS}, \\
    & m_\gamma(x) \quad \text{is SOS}, \\
    & \Re\{(\gamma - i) (\chi^{-1\dagger} - \G_{dd}^\dagger)\} \succeq \vb{0}. \label{eq:PDgammaCstrt}
\end{align}
\end{subequations}
For any valid constraint interpolation parameter $\gamma$, the inner minimization gives a bound on \eqref{eq:structOpt}; it is an SOS program over $\vx$ which we solve using the SDP solver MOSEK~\cite{mosek} combined with the SOS-to-SDP parsers in YALMIP~\cite{Lofberg2009} or SumOfSquares.py~\cite{Yuan_SumOfSquares_py}. The SOS requirement on $m_I(\vx)$ and $m_\gamma(\vx)$ is due to $C_I(\vx)$ and $C_\gamma(\vx)$ being inequality constraints which require non-negative multipliers for weak duality to hold. 
The outer minimization over $\gamma$ is to find the tightest such bound, with \eqref{eq:PDgammaCstrt} serving as a box constraint on $\gamma$ that ensures $C_\gamma(\vP)$ has a positive semi-definite quadratic form suitable for projection into the $\vx$ subspace: the projection when \eqref{eq:PDgammaCstrt} is semi-definite can be viewed as a limit of strictly definite projections. We find the optimal $\gamma$ using \textit{minimize\_scalar} in SciPy~\cite{2020SciPy-NMeth} which implements Brent's method for scalar minimization~\cite{brentAlgorithmsMinimization2002}.

\section{Results}

\begin{figure}
\centering
    \includegraphics[width=\linewidth]{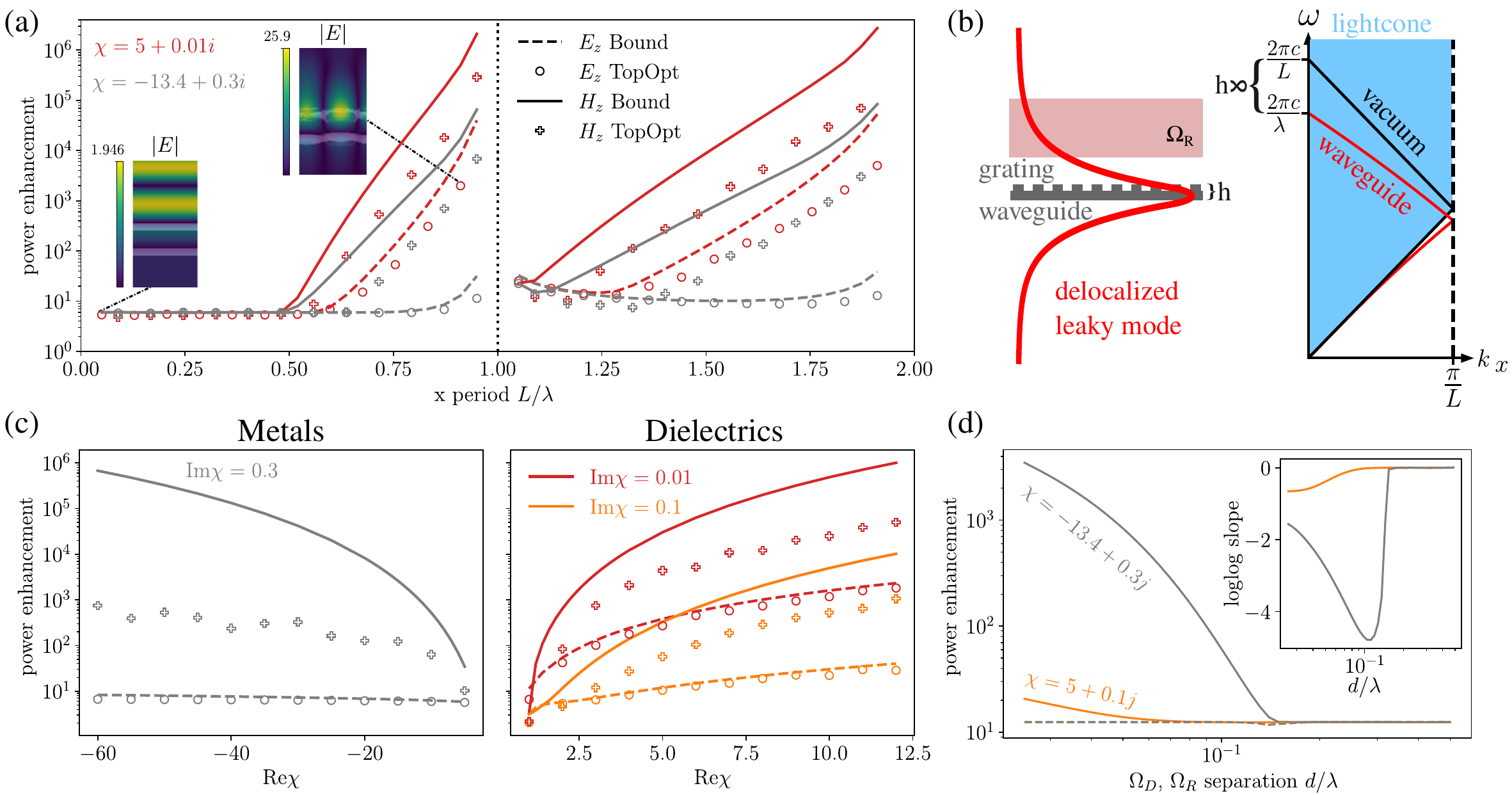}
    \caption{Bounds and inverse designs for 2D metasurfaces periodic in the $x$ direction that enhance spatially averaged Raman emission from a pump planewave traveling along the $y$ direction. Results given as factors of the SERS FOM for vacuum. (a) Varying the period $L$ for a suspended design region, $H_D = H_R = 0.5 \lambda$, separation $d=0.2\lambda$. Inset: select $E_z$ dielectric inverse design results for $L=0.125\lambda$ and $L=0.9125\lambda$. The $L=0.125\lambda$ unit-cell has been repeated 7 times for clearer viewing. (b) Band diagram and mode profile schematic for highly delocalized high $Q$ leaky modes that enable the divergence as $L$ approaches integer multiples of $\lambda$ from below. 
    (c) Varying the material susceptibility $\chi$ for a suspended design region, $L=0.8 \lambda$, $H_D = H_R = 0.5 \lambda$, separation $d=0.2\lambda$. (d) Varying $d$, for fixed $L=0.4\lambda$, $H_D=0.5\lambda$, $H_R=0.2\lambda$. Inset: the log-log slope of the $H_z$ results for comparing power law exponents of the bounds with respect to $d$. \label{fig:suspended}}
\end{figure}

To test the efficacy of our formulation, we evaluated bounds for spatially averaged SERS close to a periodic metasurface in 2D and compared with the performance of TO designed structures. Bounds were computed for suspended and substrated metasurfaces (\figref{fig:suspended} and \figref{fig:substrated} respectively) comprised of either metals or dielectrics, considering both $E_z$ (electrical field out-of-plane) and $H_z$ (electrical field in-plane) polarizations. Adjoint-gradient-based TO was done using an in-house finite difference frequency domain discretization of the structural optimization problem \eqref{eq:structOpt}. These results allow us to elucidate the scaling of maximal metasurface SERS performance with regards to key physical parameters such as the period $L$, material susceptibility $\chi$, and the molecule-metasurface separation $d$. 

\subsection{Suspended design region, vary period $L$}
\figref{fig:suspended}a shows bound and TO SERS results for suspended metasurfaces as a function of the period~$L$. The metal susceptibility $\chi=-13.4+0.3i$ corresponds to that of silver (Ag) at a wavelength $\lambda=540$ nm.  Before commenting on any specifics, we would like to first note the tightness of the bounds, which generally come within $< 10\times$ of the inverse-design performance, often within $<3\times$ for the $E_z$ polarization. Although previous upper-bounds work in linear optics has often observed similar tightness, the results here are even more impressive in light of the additional relaxations required by the quartic scaling of our objective, and is strong empirical evidence for the effectiveness of our bounding procedure. 

The maximum SERS FOM as a function of $L$ shows a distinctive plateau for $L \ll \lambda$ before exhibiting a strong divergence as $L$ approaches integer multiples of the vacuum wavelength from below (but not from above). The plateau for small $L$ suggests that the best design in that case may simply be a multilayer mirror that attempts to reflect as much of the incident planewave as possible. At larger $L$, structured metasurfaces can support guided modes with resonant enhancement of the Raman emission. The divergence as $L \rightarrow \lambda$ from below is an extreme manifestation of this: it can be achieved with thin waveguides supporting highly delocalized guided modes that couple to the incident planewave via weak gratings (\figref{fig:suspended}b).  As the waveguide becomes asymptotically thinner, such a grating resonance approaches the light line $L = \lambda$ at the edge of the Brillouin zone, with a progressively more delocalized mode (suppressing absorption loss), and a progressively weaker grating (suppressing radiation loss)~\cite{joannopoulosPhotonicCrystals2008}. The divergence is present even if the material is lossy: as $L$ increases towards $\lambda$, the thickness of the waveguide $h$ needed to support a mode with frequency $\omega_0$ also goes to 0. The mode is thus more delocalized which leads to less field concentration within $\Omega_R$ (and therefore less coupling to the Raman molecules), but this is counteracted by a simultaneous increase in $Q$ due to reduced material loss within the waveguide. On balance, the $Q$ increase dominates, and we obtain a divergence in the FOM. This motif of using delocalization to overcome loss leading to theoretically diverging performance at a single frequency has also been observed before in the context of minimizing the local density of states~\cite{strekhaSuppressingElectromagnetic2024}.  In a practical realization, however, the attainable $Q$ will be limited by fabrication disorder, finite-size effects, desired bandwidth, and similar constraints.

\subsection{Suspended design region, vary $\chi$}
\figref{fig:suspended}c illustrates the maximal SERS as a function of the material susceptibility $\chi$. For dielectrics with $\Re{\chi}>0$, the bounds and inverse design performances are proportional to $\Im{\chi}^{-2}$. Such a scaling can be accounted for by the dependence of modal field norms on $Q$ (which scales inversely with absorption $\Im \chi$), with $\int \norm{\vE}^2 \,\dd\vr \propto Q$ and hence $\int \norm{\vE}^4 \,\dd\vr \propto Q^2$.

For metals with $\Re{\chi}<0$, the most dramatic feature is the large performance gap between the $E_z$ and $H_z$ polarizations, stemming from the fact that $E_z$ polarized waves do not couple to SPPs~\cite{maierPlasmonicsFundamentals2007a}. \figref{fig:suspended}a and \figref{fig:suspended}c suggest that, away from the $L \approx \lambda$ divergences, the $E_z$ incident wave is at best fully reflected, with the FOM nearly independent of the material strength. In contrast, $H_z$ polarized waves do couple to SPPs, and we see performance trends similar to those of dielectric structures. We note that for strong metals and the $H_z$ polarization there is a bigger gap between the bounds and inverse-design performance; further investigation is needed to determine whether this is due to looseness in the bounds or the difficulties associated with metallic TO~\cite{christiansenNonlinearMaterial2019}. 

\subsection{Suspended design region, vary $d$\label{sec:resultsVaryd}}
Our ability to bound the quartic field norm also allows us to explore the theoretical maximum field-singularity strength that can be found near sharp structural features. It is well known that sharp features can support field singularities, and that (for energy conservation) these singularities must remain $L_2$ integrable ($\int \norm{\vb{E}}^2 \dd\vr$ is finite)~\cite{felsenRadiationScattering2003}. If we denote the strength of a field singularity $\norm{\vE(\vr)} \propto r^{-t}$ by the absolute value of its exponent $t$, $L_2$ integrability in 2D implies that $t<1$. Such a field singularity may still have infinite $L_4$ norm if $t>1/2$~\cite{yaoDesigningStructures2023}, which is why a finite $d$ is necessary for finite bounds on the SERS FOM. This also implies that the scaling of the SERS bounds as $d \rightarrow 0$ contains information on the maximum $t$, given that in 2D,  $\int_{r>d} \norm{\vE(\vr)}^4 \dd \vr \propto \int_{r>d} r^{-4t} \cdot r \dd r \propto d^{-(4t-2)}$. \figref{fig:suspended}d plots the bounds as a function of $d$: we see that for the $E_z$ polarization there is effectively no dependence on $d$, corroborating the fact that no singularities arise for this polarization~\cite{andersenFieldBehavior1978,budaevElectromagneticField2007}. For the $H_z$ polarization there are distinct scalings for metals and dielectrics, with the power exponents depicted in the inset. It is clear that metals can support stronger singularities than dielectrics. Additionally, both metals and dielectric bounds scale slower than $d^{-2}$, which implies a maximum $t$ substantially less than 1 for the given materials, a tightening of the $t$ upper bound given by $L_2$ integrability. Analytically deriving the exact maximum attainable scaling of field singularities dependent on $\chi$ is a challenging problem for future research; it is not as simple as optimizing over the apex angle of a solid tip, given that nanoparticle composites with fractal geometry are known to be highly effective SERS structures~\cite{moskovitsSurfaceenhancedSpectroscopy1985,wangStructuralBasis2003,liSelfSimilarChain2003}. 

\subsection{Substrated design region, vary $L$} 
\begin{figure}
    \centering
    \includegraphics[width=0.75\linewidth]{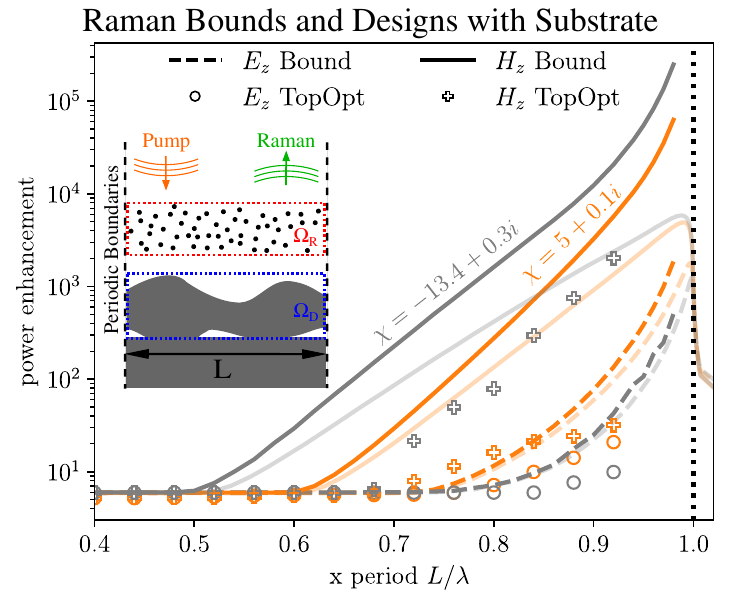}
    \caption{Bounds (lines) and inverse designs (shapes) for substrated metasurfaces (inset schematic) as a function of the period $L$ with other geometry parameters $H_D = H_R=0.5\lambda$, $d=0.2\lambda$. Results given as factors of the SERS FOM for a bare substrate. 
    Translucent lines are $H_z$ bounds given a uniform loss factor applied across space: $\epsilon(\vr) \rightarrow (1+\frac{i}{600})\epsilon(\vr)$, to cap the maximum modal $Q$.} 
    \label{fig:substrated}
\end{figure}

From a fabrication point of view, it is often more convenient to fabricate a metasurface on top of bulk material instead of suspending it. We investigate this scenario by considering bounds and inverse designs given an extended substrate below $\Omega_D$ with the same susceptibility $\chi$; the results are shown in \figref{fig:substrated}. For small $L$, we again see a convergence of performance between different materials and polarizations to a fixed value corresponding to multilayer designs that optimally reflect the incident planewave. As $L$ increases towards $\lambda$, the presence of a lossy substrate rules out the delocalized guided mode strategy that led to divergences in \figref{fig:suspended}a. However, a divergence in the bounds still persists, albeit with a reduced logarithmic dependence $\propto |\log(\lambda - L)|$ compared to $\propto 1/(\lambda-L)^2$ for suspended structures (SI); for $H_z$ polarized metallic structures the inverse designs still track the bounds closely. A physical explanation of the origin of this logarithmic divergence is left for future work. To regularize such divergences, we can multiply $\epsilon(\vr)$ across space by a uniform loss factor, concretely taken to be $1+i/600$, which physically has the effect of capping the quality factor of any given mode to $\lesssim 300$~\cite{liangFormulationScalable2013} and is thus more relevant to practical settings where finite bandwidth requirements and fabrication imperfections need to be accounted for. The $Q$-capped bounds do not diverge but rather display a sharp finite transition around $L=\lambda$. 

\section{Summary and Outlook}

To summarize, we presented the first bounds on spatially averaged SERS derived using SOS programming. The bounds generally agree with the performance of inverse designs, and provide detailed information on the performance scaling with regards to key physical parameters such as the design region geometry and material susceptibility. Metasurfaces supporting high $Q$ guided modes are shown to be an optimal strategy for maximizing large area SERS, with some techniques for reducing absorption loss by suspending the structure and making use of more delocalized modes. Furthermore, SOS bounding of the quartic field norm FOM offers a new way of investigating properties of field singularities in Maxwell's equations. 

Going forward, there are multiple avenues for further investigation. Given the additional complexity of SOS programming, we have chosen to compute the bounds with just global power conservation constraints over $\Omega_D$ for simplicity. The inclusion of more localized constraints has proven effective in tightening bounds for linear photonics design~\cite{moleskyHierarchicalMeanfield2020,kuangComputationalBounds2020}, and should also be useful here. The incorporation of more constraints into the SOS framework may necessitate further numerical efficiency improvements, possibly by using weaker SOS formulations based on linear or second-order cone programming~\cite{ahmadiDSOSSDSOS2019}. Finally, the combination of dimension reduction techniques (such as low-rank Green's function approximations) with SOS programming is a promising direction of study for other inverse design problems involving non-linear phenomenon in photonics and other branches of wave physics. The exact form of the nonlinearity is not critical, so long as an appropriate FOM can be formulated as a polynomial (or rational function~\cite{angerisBoundsEfficiency2023}) of the fields. Specific problems of interest include minimizing the bistability threshold for Kerr-nonlinearity-based optical switching~\cite{almeidaOpticalBistability2004,wangOpticalBistability2008,daiLowThreshold2015}, improving the frequency conversion efficiency of nonlinear metasurfaces~\cite{gwoPlasmonicMetasurfaces2016,krasnokNonlinearMetasurfaces2018,huangPlanarNonlinear2020}, enhanced sensitivity of nonlinear optical detectors~\cite{petersExceptionalPrecision2022,sunBayesianOptimization2024}, and reducing the lasing threshold of nanostructured semiconductor lasers~\cite{ellisUltralowthresholdElectrically2011,wuMonolayerSemiconductor2015,diezInverseDesign2023}. 

\section{Back matter}
\begin{backmatter}

\bmsection{Funding} This work was supported in part by the Simons Foundation, and by the U.S. Army Research Office (ARO) through the Institute for Soldier Nanotechnologies (ISN) under award no. W911NF-18-2-0048.

\bmsection{Acknowledgment}
We would like to thank Alejandro Rodriguez, Sean Molesky, Alessio Amaolo, Luiz Faria, and Carlos Perez Arancibia for helpful discussions. We are grateful to the MIT SuperCloud and Lincoln Laboratory Supercomputing Center for providing HPC resources that have contributed to the research results reported within this paper.

\bmsection{Disclosures}
The authors declare no conflicts of interest.

\bmsection{Data Availability Statement}
Data underlying the results presented in this paper are not publicly available at this time but may be obtained from the authors upon reasonable request.

\bmsection{Supplemental document}
See supplementary info for supporting content. 

\end{backmatter}

\bibliography{raman_refs.bib,solvers.bib}

\begin{thebibliography}{10}
\newcommand{\enquote}[1]{``#1''}

\bibitem{leruRigorousJustification2006}
E.~C. Le~Ru and P.~G. Etchegoin, \enquote{Rigorous justification of the {\textbar}{{{\emph{E}}}}{\textbar}4 enhancement factor in {{Surface Enhanced Raman Spectroscopy}},} {\protect\JournalTitle{Chemical Physics Letters}} \textbf{423}, 63--66 (2006).

\bibitem{yaoDesigningStructures2023}
W.~Yao, F.~Verdugo, H.~O. Everitt, \emph{et~al.}, \enquote{Designing structures that maximize spatially averaged surface-enhanced {{Raman}} spectra,} {\protect\JournalTitle{Optics Express}} \textbf{31}, 4964--4977 (2023).

\bibitem{parriloStructuredSemidefinite2000}
P.~A. Parrilo, \enquote{Structured semidefinite programs and semialgebraic geometry methods in robustness and optimization,} Ph.D. thesis, California Institute of Technology (2000).

\bibitem{lasserreGlobalOptimization2001}
J.~B. Lasserre, \enquote{Global {{Optimization}} with {{Polynomials}} and the {{Problem}} of {{Moments}},} {\protect\JournalTitle{SIAM Journal on Optimization}} \textbf{11}, 796--817 (2001).

\bibitem{blekhermanSemidefiniteOptimization2013}
G.~Blekherman, P.~A. Parrilo, and R.~R. Thomas, eds., \emph{Semidefinite Optimization and Convex Algebraic Geometry}, {{MOS-SIAM}} Series on Optimization ({Society for Industrial and Applied Mathematics : Mathematical Programming Society}, Philadelphia, 2013).

\bibitem{chaoPhysicalLimits2022}
P.~Chao, B.~Strekha, R.~Kuate~Defo, \emph{et~al.}, \enquote{Physical limits in electromagnetism,} {\protect\JournalTitle{Nature Reviews Physics}} \textbf{4}, 543--559 (2022).

\bibitem{angerisHeuristicMethods2021}
G.~Angeris, J.~Vu{\v c}kovi{\'c}, and S.~Boyd, \enquote{Heuristic methods and performance bounds for photonic design,} {\protect\JournalTitle{Optics Express}} \textbf{29}, 2827--2854 (2021).

\bibitem{gertlerManyPhotonic2025}
S.~Gertler, Z.~Kuang, C.~Christie, \emph{et~al.}, \enquote{Many photonic design problems are sparse {{QCQPs}},} {\protect\JournalTitle{Science Advances}}  (2025).

\bibitem{strekhaSuppressingElectromagnetic2024}
B.~Strekha, P.~Chao, R.~K. Defo, \emph{et~al.}, \enquote{Suppressing electromagnetic local density of states via slow light in lossy quasi-one-dimensional gratings,} {\protect\JournalTitle{Physical Review A}} \textbf{109}, L041501 (2024).

\bibitem{maierPlasmonicsFundamentals2007a}
S.~A. Maier, \emph{Plasmonics: {{Fundamentals}} and {{Applications}}} (Springer US, New York, NY, 2007).

\bibitem{felsenRadiationScattering2003}
L.~B. Felsen and N.~Marcuvitz, \emph{Radiation and Scattering of Waves}, {{IEEE Press}} Series on Electromagnetic Waves ({Inst. of Electrical and Electronics Engineers}, New York, 2003).

\bibitem{idemenConfluentTip2003}
M.~Idemen, \enquote{Confluent tip singularity of the electromagnetic field at the apex of a material cone,} {\protect\JournalTitle{Wave Motion}} \textbf{38}, 251--277 (2003).

\bibitem{budaevElectromagneticField2007}
B.~V. Budaev and D.~B. Bogy, \enquote{On the electromagnetic field singularities near the vertex of a dielectric wedge,} {\protect\JournalTitle{Radio Science}} \textbf{42} (2007).

\bibitem{kottmannSpectralResponse2000}
J.~P. Kottmann, O.~J.~F. Martin, D.~R. Smith, and S.~Schultz, \enquote{Spectral response of plasmon resonant nanoparticles with a non-regular shape,} {\protect\JournalTitle{Optics Express}} \textbf{6}, 213--219 (2000).

\bibitem{garrellSurfaceenhancedRaman1989}
R.~L. Garrell, \enquote{Surface-enhanced {{Raman}} spectroscopy,} {\protect\JournalTitle{Analytical Chemistry}} \textbf{61}, 401A--411A (1989).

\bibitem{kneippSurfaceEnhancedRaman2006}
K.~Kneipp, M.~Moskovits, and H.~Kneipp, eds., \emph{Surface-{{Enhanced Raman Scattering}}}, vol. 103 of \emph{Topics in {{Applied Physics}}} (Springer Berlin Heidelberg, 2006).

\bibitem{jonesRamanTechniques2019}
R.~R. Jones, D.~C. Hooper, L.~Zhang, \emph{et~al.}, \enquote{Raman {{Techniques}}: {{Fundamentals}} and {{Frontiers}},} {\protect\JournalTitle{Nanoscale Research Letters}} \textbf{14}, 231 (2019).

\bibitem{moskovitsSurfaceenhancedSpectroscopy1985}
M.~Moskovits, \enquote{Surface-enhanced spectroscopy,} {\protect\JournalTitle{Reviews of Modern Physics}} \textbf{57}, 783--826 (1985).

\bibitem{langerPresentFuture2020}
J.~Langer, D.~{Jimenez de Aberasturi}, J.~Aizpurua, \emph{et~al.}, \enquote{Present and {{Future}} of {{Surface-Enhanced Raman Scattering}},} {\protect\JournalTitle{ACS Nano}} \textbf{14}, 28--117 (2020).

\bibitem{moleskyInverseDesign2018}
S.~Molesky, Z.~Lin, A.~Y. Piggott, \emph{et~al.}, \enquote{Inverse design in nanophotonics,} {\protect\JournalTitle{Nature Photonics}} \textbf{12}, 659--670 (2018).

\bibitem{christiansenInverseDesign2020}
R.~E. Christiansen, J.~Michon, M.~Benzaouia, \emph{et~al.}, \enquote{Inverse design of nanoparticles for enhanced {{Raman}} scattering,} {\protect\JournalTitle{Optics Express}} \textbf{28}, 4444--4462 (2020).

\bibitem{panTopologyOptimization2021}
Y.~Pan, R.~E. Christiansen, J.~Michon, \emph{et~al.}, \enquote{Topology optimization of surface-enhanced {{Raman}} scattering substrates,} {\protect\JournalTitle{Applied Physics Letters}} \textbf{119}, 061601 (2021).

\bibitem{andersenFieldBehavior1978}
J.~Andersen and V.~Solodukhov, \enquote{Field behavior near a dielectric wedge,} {\protect\JournalTitle{IEEE Transactions on Antennas and Propagation}} \textbf{26}, 598--602 (1978).

\bibitem{mcmahonNonlocalOptical2009}
J.~M. McMahon, S.~K. Gray, and G.~C. Schatz, \enquote{Nonlocal {{Optical Response}} of {{Metal Nanostructures}} with {{Arbitrary Shape}},} {\protect\JournalTitle{Physical Review Letters}} \textbf{103}, 097403 (2009).

\bibitem{ciraciProbingUltimate2012}
C.~Cirac{\`i}, R.~T. Hill, J.~J. Mock, \emph{et~al.}, \enquote{Probing the {{Ultimate Limits}} of {{Plasmonic Enhancement}},} {\protect\JournalTitle{Science}} \textbf{337}, 1072--1074 (2012).

\bibitem{gubbinOpticalNonlocality2020}
C.~R. Gubbin and S.~De~Liberato, \enquote{Optical {{Nonlocality}} in {{Polar Dielectrics}},} {\protect\JournalTitle{Physical Review X}} \textbf{10}, 021027 (2020).

\bibitem{zhouMinimumLength2015a}
M.~Zhou, B.~S. Lazarov, F.~Wang, and O.~Sigmund, \enquote{Minimum length scale in topology optimization by geometric constraints,} {\protect\JournalTitle{Computer Methods in Applied Mechanics and Engineering}} \textbf{293}, 266--282 (2015).

\bibitem{vercruysseAnalyticalLevel2019a}
D.~Vercruysse, N.~V. Sapra, L.~Su, \emph{et~al.}, \enquote{Analytical level set fabrication constraints for inverse design,} {\protect\JournalTitle{Scientific Reports}} \textbf{9}, 8999 (2019).

\bibitem{hammondPhotonicTopology2021}
A.~M. Hammond, A.~Oskooi, S.~G. Johnson, and S.~E. Ralph, \enquote{Photonic topology optimization with semiconductor-foundry design-rule constraints,} {\protect\JournalTitle{Optics Express}} \textbf{29}, 23916 (2021).

\bibitem{schubertInverseDesign2022}
M.~F. Schubert, A.~K.~C. Cheung, I.~A.~D. Williamson, \emph{et~al.}, \enquote{Inverse {{Design}} of {{Photonic Devices}} with {{Strict Foundry Fabrication Constraints}},} {\protect\JournalTitle{ACS Photonics}} \textbf{9}, 2327--2336 (2022).

\bibitem{hackbuschHierarchicalMatrices2015}
W.~Hackbusch, \emph{Hierarchical {{Matrices}}: {{Algorithms}} and {{Analysis}}}, vol.~49 of \emph{Springer {{Series}} in {{Computational Mathematics}}} (Springer, Berlin, Heidelberg, 2015).

\bibitem{polimeridisFluctuatingVolumecurrent2015}
A.~G. Polimeridis, M.~T.~H. Reid, W.~Jin, \emph{et~al.}, \enquote{Fluctuating volume-current formulation of electromagnetic fluctuations in inhomogeneous media: {{Incandescence}} and luminescence in arbitrary geometries,} {\protect\JournalTitle{Physical Review B}} \textbf{92}, 134202 (2015).

\bibitem{chaoMaximumElectromagnetic2022}
P.~Chao, R.~K. Defo, S.~Molesky, and A.~Rodriguez, \enquote{Maximum electromagnetic local density of states via material structuring,} {\protect\JournalTitle{Nanophotonics}}  (2022).

\bibitem{gustafssonUpperBounds2020}
M.~Gustafsson, K.~Schab, L.~Jelinek, and M.~Capek, \enquote{Upper bounds on absorption and scattering,} {\protect\JournalTitle{New Journal of Physics}} \textbf{22}, 073013 (2020).

\bibitem{moleskyGlobal$mathbbT$2020}
S.~Molesky, P.~Chao, W.~Jin, and A.~W. Rodriguez, \enquote{Global {$\mathbb{T}$}-operator bounds on electromagnetic scattering: Upper bounds on far-field cross sections,} {\protect\JournalTitle{Physical Review Research}} \textbf{2}, 033172 (2020).

\bibitem{kuangComputationalBounds2020}
Z.~Kuang and O.~D. Miller, \enquote{Computational bounds to light--matter interactions via local conservation laws,} {\protect\JournalTitle{Physical Review Letters}} \textbf{125}, 263607 (2020).

\bibitem{shimFundamentalLimits2019}
H.~Shim, L.~Fan, S.~G. Johnson, and O.~D. Miller, \enquote{Fundamental {{Limits}} to {{Near-Field Optical Response}} over {{Any Bandwidth}},} {\protect\JournalTitle{Physical Review X}} \textbf{9}, 011043 (2019).

\bibitem{molesky$mathbbT$OperatorLimits2021}
S.~Molesky, P.~Chao, J.~Mohajan, \emph{et~al.}, \enquote{{$\mathbb{T}$}-operator limits on optical communication: {{Metaoptics}}, computation, and input-output transformations,} {\protect\JournalTitle{Physical Review Research}} \textbf{4}, 013020 (2022).

\bibitem{shimFundamentalLimits2021}
H.~Shim, Z.~Kuang, Z.~Lin, and O.~D. Miller, \enquote{Fundamental limits to multi-functional and tunable nanophotonic response,}  (2021).

\bibitem{venkataramFundamentalLimits2020}
P.~S. Venkataram, S.~Molesky, P.~Chao, and A.~W. Rodriguez, \enquote{Fundamental limits to attractive and repulsive {{Casimir-Polder}} forces,} {\protect\JournalTitle{Physical Review A}} \textbf{101}, 052115 (2020).

\bibitem{strekhaTraceExpressions2024}
B.~Strekha, M.~Kr{\"u}ger, and A.~W. Rodriguez, \enquote{Trace expressions and associated limits for equilibrium {{Casimir}} torque,} {\protect\JournalTitle{Physical Review A}} \textbf{109}, 012813 (2024).

\bibitem{mohajanFundamentalLimits2023}
J.~Mohajan, P.~Chao, W.~Jin, \emph{et~al.}, \enquote{Fundamental limits on radiative {$\chi$}{\textsuperscript{(2)}} second harmonic generation,} {\protect\JournalTitle{Optics Express}} \textbf{31}, 44212--44223 (2023).

\bibitem{michonLimitsSurfaceenhanced2019}
J.~Michon, M.~Benzaouia, W.~Yao, \emph{et~al.}, \enquote{Limits to surface-enhanced {{Raman}} scattering near arbitrary-shape scatterers,} {\protect\JournalTitle{Optics Express}} \textbf{27}, 35189--35202 (2019).

\bibitem{amaoloPhysicalLimits2024}
A.~Amaolo, P.~Chao, T.~J. Maldonado, \emph{et~al.}, \enquote{Physical limits on {{Raman}} scattering: {{The}} critical role of pump and signal co-design,} {\protect\JournalTitle{Physical Review A}} \textbf{110}, L061501 (2024).

\bibitem{linCavityenhancedSecondharmonic2016}
Z.~Lin, X.~Liang, M.~Lon{\v c}ar, \emph{et~al.}, \enquote{Cavity-enhanced second-harmonic generation via nonlinear-overlap optimization,} {\protect\JournalTitle{Optica}} \textbf{3}, 233--238 (2016).

\bibitem{sitawarinInversedesignedPhotonic2018a}
C.~Sitawarin, W.~Jin, Z.~Lin, and A.~W. Rodriguez, \enquote{Inverse-designed photonic fibers and metasurfaces for nonlinear frequency conversion [{{Invited}}],} {\protect\JournalTitle{Photonics Research}} \textbf{6}, B82--B89 (2018).

\bibitem{hughesAdjointMethod2018}
T.~W. Hughes, M.~Minkov, I.~A.~D. Williamson, and S.~Fan, \enquote{Adjoint {{Method}} and {{Inverse Design}} for {{Nonlinear Nanophotonic Devices}},} {\protect\JournalTitle{ACS Photonics}} \textbf{5}, 4781--4787 (2018).

\bibitem{yangInversedesignedSilicon2023}
J.~Yang, M.~A. Guidry, D.~M. Lukin, \emph{et~al.}, \enquote{Inverse-designed silicon carbide quantum and nonlinear photonics,} {\protect\JournalTitle{Light: Science \& Applications}} \textbf{12}, 201 (2023).

\bibitem{mannInverseDesign2023a}
S.~A. Mann, H.~Goh, and A.~Al{\`u}, \enquote{Inverse {{Design}} of {{Nonlinear Polaritonic Metasurfaces}} for {{Second Harmonic Generation}},} {\protect\JournalTitle{ACS Photonics}} \textbf{10}, 993--1000 (2023).

\bibitem{stichInverseDesign2024}
S.~Stich, J.~Mohajan, D.~de~Ceglia, \emph{et~al.}, \enquote{Inverse {{Design}} of an {{All-Dielectric Nonlinear Polaritonic Metasurface}},}  (2024).

\bibitem{hammondDesigningStructures2024}
I.~M. Hammond, P.~Chao, W.~Yao, \emph{et~al.}, \enquote{Designing structures that maximize spatially averaged surface-enhanced {{Raman}} spectra: Erratum,} {\protect\JournalTitle{Optics Express}} \textbf{32}, 44754--44755 (2024).

\bibitem{parkGeneralHeuristics2017a}
J.~Park and S.~Boyd, \enquote{General {{Heuristics}} for {{Nonconvex Quadratically Constrained Quadratic Programming}},}  (2017).

\bibitem{boydConvexOptimization2004}
S.~P. Boyd and L.~Vandenberghe, \emph{Convex Optimization} (Cambridge University Press, Cambridge, UK ; New York, 2004).

\bibitem{ahmadiNPhardnessDeciding2013}
A.~A. Ahmadi, A.~Olshevsky, P.~A. Parrilo, and J.~N. Tsitsiklis, \enquote{{{NP-hardness}} of deciding convexity of quartic polynomials and related problems,} {\protect\JournalTitle{Mathematical Programming}} \textbf{137}, 453--476 (2013).

\bibitem{parriloMinimizingPolynomial2001}
P.~A. Parrilo and B.~Sturmfels, \enquote{Minimizing {{Polynomial Functions}},}  (2001).

\bibitem{ahmadiDSOSSDSOS2019}
A.~A. Ahmadi and A.~Majumdar, \enquote{{{DSOS}} and {{SDSOS Optimization}}: {{More Tractable Alternatives}} to {{Sum}} of {{Squares}} and {{Semidefinite Optimization}},} {\protect\JournalTitle{SIAM Journal on Applied Algebra and Geometry}} \textbf{3}, 193--230 (2019).

\bibitem{mosek}
M.~ApS, \emph{MOSEK optimizer API for python manual. Release 10.2.13} (2025).

\bibitem{Lofberg2009}
J.~L{\"{o}}fberg, \enquote{Pre- and post-processing sum-of-squares programs in practice,} {\protect\JournalTitle{IEEE Transactions on Automatic Control}} \textbf{54}, 1007--1011 (2009).

\bibitem{Yuan_SumOfSquares_py}
C.~Yuan, \enquote{{SumOfSquares.py},} .

\bibitem{2020SciPy-NMeth}
P.~Virtanen, R.~Gommers, T.~E. Oliphant, \emph{et~al.}, \enquote{{{SciPy} 1.0: Fundamental Algorithms for Scientific Computing in Python},} {\protect\JournalTitle{Nature Methods}} \textbf{17}, 261--272 (2020).

\bibitem{brentAlgorithmsMinimization2002}
R.~P. Brent, \emph{Algorithms for Minimization without Derivatives} (Dover Publications, Mineola, N.Y, 2002).

\bibitem{joannopoulosPhotonicCrystals2008}
J.~D. Joannopoulos, J.~G. Steven, J.~N. Winn, and R.~D. Meade, \emph{Photonic Crystals: Molding the Flow of Light} (Princeton University Press, Princeton, 2008), 2nd ed.

\bibitem{christiansenNonlinearMaterial2019}
R.~E. Christiansen, J.~{Vester-Petersen}, S.~P. Madsen, and O.~Sigmund, \enquote{A non-linear material interpolation for design of metallic nano-particles using topology optimization,} {\protect\JournalTitle{Computer Methods in Applied Mechanics and Engineering}} \textbf{343}, 23--39 (2019).

\bibitem{wangStructuralBasis2003}
Z.~Wang, S.~Pan, T.~D. Krauss, \emph{et~al.}, \enquote{The structural basis for giant enhancement enabling single-molecule {{Raman}} scattering,} {\protect\JournalTitle{Proceedings of the National Academy of Sciences}} \textbf{100}, 8638--8643 (2003).

\bibitem{liSelfSimilarChain2003}
K.~Li, M.~I. Stockman, and D.~J. Bergman, \enquote{Self-{{Similar Chain}} of {{Metal Nanospheres}} as an {{Efficient Nanolens}},} {\protect\JournalTitle{Physical Review Letters}} \textbf{91}, 227402 (2003).

\bibitem{liangFormulationScalable2013}
X.~Liang and S.~G. Johnson, \enquote{Formulation for scalable optimization of microcavities via the frequency-averaged local density of states,} {\protect\JournalTitle{Optics Express}} \textbf{21}, 30812--30841 (2013).

\bibitem{moleskyHierarchicalMeanfield2020}
S.~Molesky, P.~Chao, and A.~W. Rodriguez, \enquote{Hierarchical mean-field {{T-operator}} bounds on electromagnetic scattering: {{Upper}} bounds on near-field radiative {{Purcell}} enhancement,} {\protect\JournalTitle{Physical Review Research}} \textbf{2}, 043398 (2020).

\bibitem{angerisBoundsEfficiency2023}
G.~Angeris, T.~Diamandis, J.~Vu{\v c}kovi{\'c}, and S.~P. Boyd, \enquote{Bounds on {{Efficiency Metrics}} in {{Photonics}},} {\protect\JournalTitle{ACS Photonics}} \textbf{10}, 2521--2529 (2023).

\bibitem{almeidaOpticalBistability2004}
V.~R. Almeida and M.~Lipson, \enquote{Optical bistability on a silicon chip,} {\protect\JournalTitle{Optics Letters}} \textbf{29}, 2387--2389 (2004).

\bibitem{wangOpticalBistability2008}
F.~Y. Wang, G.~X. Li, H.~L. Tam, \emph{et~al.}, \enquote{Optical bistability and multistability in one-dimensional periodic metal-dielectric photonic crystal,} {\protect\JournalTitle{Applied Physics Letters}} \textbf{92}, 211109 (2008).

\bibitem{daiLowThreshold2015}
X.~Dai, L.~Jiang, and Y.~Xiang, \enquote{Low threshold optical bistability at terahertz frequencies with graphene surface plasmons,} {\protect\JournalTitle{Scientific Reports}} \textbf{5}, 12271 (2015).

\bibitem{gwoPlasmonicMetasurfaces2016}
S.~Gwo, C.-Y. Wang, H.-Y. Chen, \emph{et~al.}, \enquote{Plasmonic {{Metasurfaces}} for {{Nonlinear Optics}} and {{Quantitative SERS}},} {\protect\JournalTitle{ACS Photonics}} \textbf{3}, 1371--1384 (2016).

\bibitem{krasnokNonlinearMetasurfaces2018}
A.~Krasnok, M.~Tymchenko, and A.~Al{\`u}, \enquote{Nonlinear metasurfaces: A paradigm shift in nonlinear optics,} {\protect\JournalTitle{Materials Today}} \textbf{21}, 8--21 (2018).

\bibitem{huangPlanarNonlinear2020}
T.~Huang, X.~Zhao, S.~Zeng, \emph{et~al.}, \enquote{Planar nonlinear metasurface optics and their applications,} {\protect\JournalTitle{Reports on Progress in Physics}} \textbf{83}, 126101 (2020).

\bibitem{petersExceptionalPrecision2022}
K.~J.~H. Peters and S.~R.~K. Rodriguez, \enquote{Exceptional {{Precision}} of a {{Nonlinear Optical Sensor}} at a {{Square-Root Singularity}},} {\protect\JournalTitle{Physical Review Letters}} \textbf{129}, 013901 (2022).

\bibitem{sunBayesianOptimization2024}
M.~Sun, V.~Kovanis, M.~Lon{\v c}ar, and Z.~Lin, \enquote{Bayesian optimization of {{Fisher Information}} in nonlinear multiresonant quantum photonics gyroscopes,} {\protect\JournalTitle{Nanophotonics}} \textbf{13}, 2401--2416 (2024).

\bibitem{ellisUltralowthresholdElectrically2011}
B.~Ellis, M.~A. Mayer, G.~Shambat, \emph{et~al.}, \enquote{Ultralow-threshold electrically pumped quantum-dot photonic-crystal nanocavity laser,} {\protect\JournalTitle{Nature Photonics}} \textbf{5}, 297--300 (2011).

\bibitem{wuMonolayerSemiconductor2015}
S.~Wu, S.~Buckley, J.~R. Schaibley, \emph{et~al.}, \enquote{Monolayer semiconductor nanocavity lasers with ultralow thresholds,} {\protect\JournalTitle{Nature}} \textbf{520}, 69--72 (2015).

\bibitem{diezInverseDesign2023}
I.~Diez, A.~Krysa, and I.~J. Luxmoore, \enquote{Inverse {{Design}} of {{Whispering-Gallery Nanolasers}} with {{Tailored Beam Shape}} and {{Polarization}},} {\protect\JournalTitle{ACS Photonics}} \textbf{10}, 968--976 (2023).

\end{thebibliography}

\end{document}

\title{Sum-of-Squares Bounds on Surface-Enhanced Raman Scattering: Supplementary Info}

\author{Pengning Chao, Ian M. Hammond, and Steven G. Johnson}
\date{}
\maketitle

\section{Low-dimensional projection of high-dimensional definite quadratic constraints}
\begin{figure}
\centering
    \includegraphics[width=0.5\linewidth]{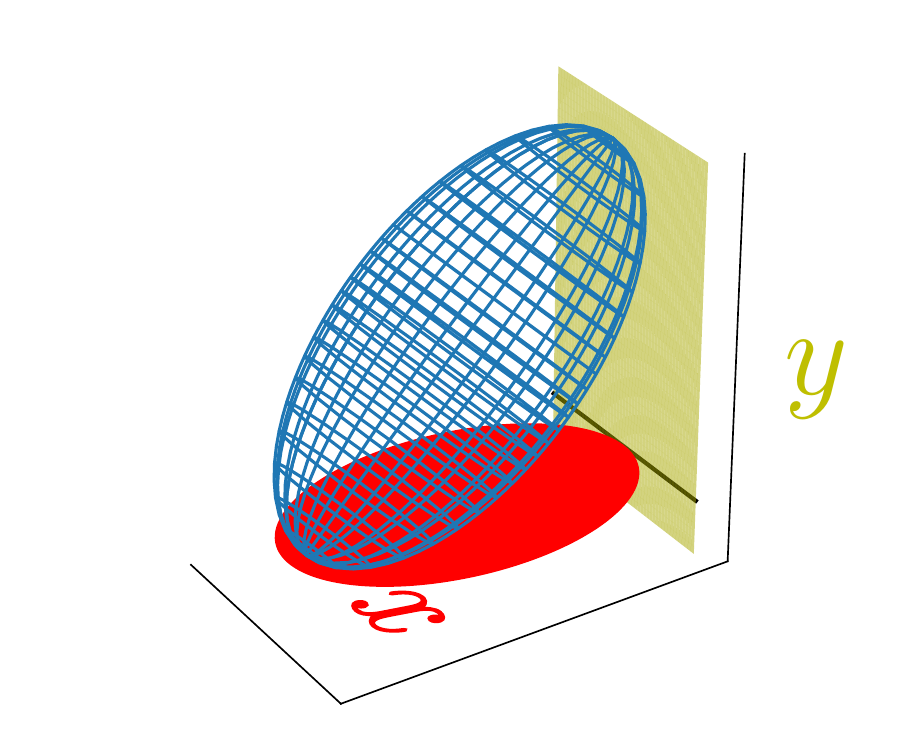}
    \caption{\label{fig:ellipsoidProj} Schematic of the ellipsoidal surface projection: the ellipsoidal surface $f(x,y)=0$ (blue) is projected onto the $x$ subspace, giving a filled-in ellipse (red). The boundaries of the projected ellipse are projected from $(x,y)$ points where the tangent hyperplane to $f(x,y)=0$ has a normal vector parallel to the $x$ subspace. }
\end{figure}
Consider a quadratic equality constraint on some high dimensional complex vector $p \in \mathbb{C}^n$ of the form
\begin{equation}
    \Re{u^\dagger p} - p^\dagger H p = 0, 
    \label{eq:quadpH}
\end{equation}
where the Hermitian matrix $H$ is positive definite. In our context, $p$ is the polarization field within the design region under some basis, and the constraint may be the global conservation of resistive power or some positive definite linear combination of the resistive and reactive power constraints. Since we focus on just the part of $p$ within the large singular vector subspace of $\G_{RD}$ in evaluating the SOS bounds, we would like to project (\ref{eq:quadpH}) onto this subspace. Given $H \succ 0$, (\ref{eq:quadpH}) specifies an ellipsoidal surface in $\mathbb{C}^n$, and the projection would be a filled-in ellipsoid representing an inequality constraint in the large singular vector subspace; see Figure~\ref{fig:ellipsoidProj} for a schematic.

Let us assume that a suitable basis transformation has been done so (\ref{eq:quadpH}) is given in the right singular vector basis of $\G_{RD}$, and write (\ref{eq:quadpH}) in block form:
\begin{equation}
    f(x,y) \equiv \Re{\mqty[u_{x} \\ u_{y}]^\dagger \mqty[x \\ y]} - \mqty[x\\y]^\dagger \mqty[J & L \\ L^\dagger & K] \mqty[x\\y] = 0
    \label{eq:quadxy}
\end{equation}
where $x$ gives the coefficients of $p$ in the large singular vector subspace and $y$ are the remaining coefficients in the smaller singular vectors. We seek an algebraic representation of the projection given by
\begin{equation}
     - x^\dagger A x + \Re{v^\dagger x} + c \geq 0.
    \label{eq:quadxA}
\end{equation}
To do so, we make use of the fact that the boundary of (\ref{eq:quadxA}) is given by the projection of points $(x,y)$ on (\ref{eq:quadxy}) such that the surface normal vector $\grad f(x,y)$ lies completely within the $x$ subspace. Using CR calculus, we have
\begin{equation}
    \pdv{f}{p^\dagger} = \frac{u}{2} - H p
\end{equation}
which implies that these boundary projection points satisfy
\begin{equation}
    \frac{u_y}{2} - (L^\dagger x + Ky) = 0.
    \label{eq:fgrad_parallel}
\end{equation}
(\ref{eq:fgrad_parallel}) can now be used to eliminate $y$ in (\ref{eq:quadxy}), giving an equation for the boundary of (\ref{eq:quadxA})
\begin{equation}
    -x^\dagger (J - LK^{-1}L^\dagger)x + \Re{(u_x - LK^{-1}u_y)^\dagger x} + \frac{1}{4}u_y^\dagger K^{-1} u_y = 0
\end{equation}
allowing us to conclude that
\begin{align}
    A &= J - LK^{-1}L^\dagger, \\
    v &= u_x - LK^{-1}u_y, \\
    c &= \frac{1}{4}u_y^\dagger K^{-1} u_y.
\end{align}
Note that $A$ is the Schur complement of the $y$ block of $H$, and is positive definite given that $H$ is.

\section{Divergence of bounds as $L$ approaches $\lambda$}
\begin{figure}[htp!]
    \centering
    \includegraphics[width=\linewidth]{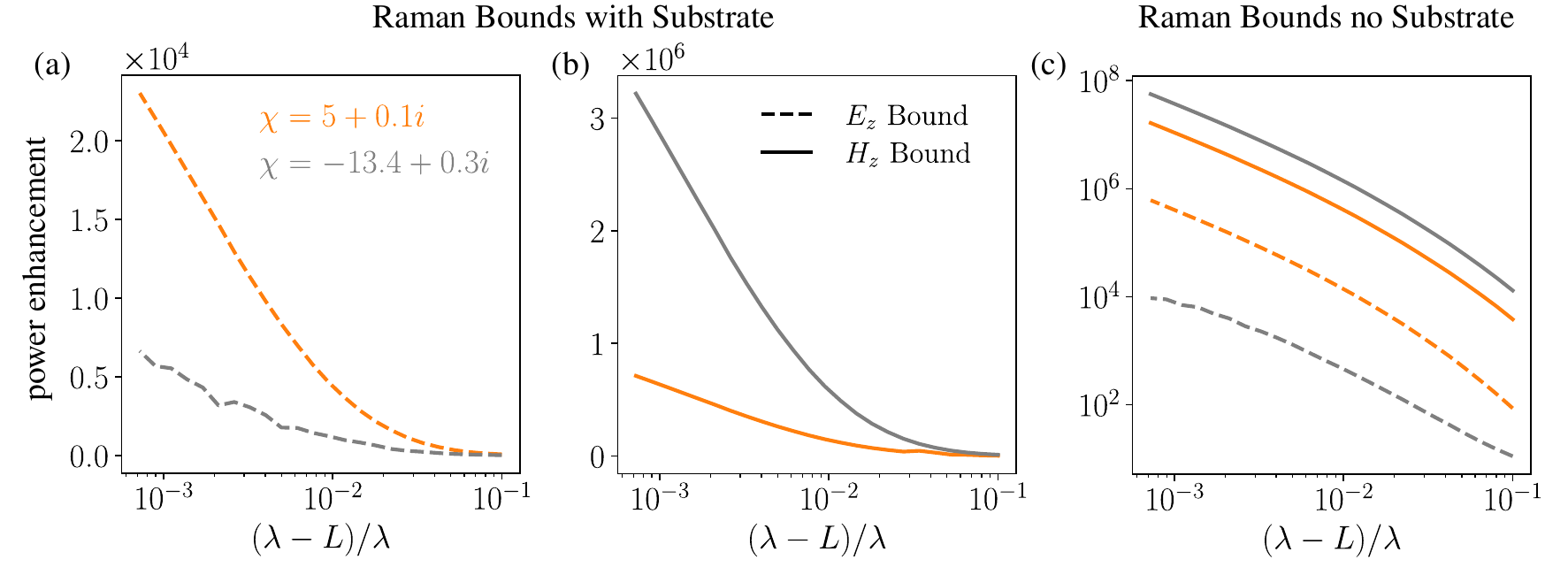}
    \caption{SERS bounds given (params here) as $L \rightarrow \lambda$ from below, for both substrated (a) and suspended (b) design regions. (a) has linear scales for the power enhancement FOM and log scales for the separation $\lambda - L$, demonstrating a logarithmic divergence FOM$\,\sim \log(\lambda-L)$ as $L \rightarrow \lambda^-$ for substrated design regions. (b) has a log-log scale and shows that FOM$\,\sim (\lambda-L)^{-2}$ as $L \rightarrow \lambda^-$ for suspended design regions. \label{fig:LcloseTo1}}
\end{figure}

To better illustrate the divergences of the periodic metasurface SERS bounds as the unit cell size $L$ approaches an integer multiple of the vacuum wavelength $\lambda$ from below, Figure S\ref{fig:LcloseTo1} plots the SERS FOM enhancement as a function of $(\lambda-L)/\lambda$ as $L \rightarrow \lambda^-$. 

\section{Runtime Benchmarking}
As discussed in the Formulation section of the main text, to facilitate computation of the SOS bounds, we represent the polarization $\vP$ using the right singular vectors $\{\vb{v}_j\}$ of the vacuum Green's function $\G_{RD} = \sum_{j=1}^\infty s_j \vb{u}_j \vb{v}_j^\dagger$. The quartic field norm FOM can then be split into a dominant singular vector subspace contribution and a remainder, given a mode cutoff $q$:
\begin{equation}
    f(\vP) \leq \left( \norm{\vE_i + \sum_{j=1}^q s_j (\vb{v}_j^\dagger \vP) \vb{u}_j}_4 + \norm{\sum_{j=q+1}^\infty s_j (\vb{v}_j^\dagger \vP) \vb{u}_j}_4  \right)^4. \numthis \label{SIeq:L4triangle}
\end{equation}
We bound the first term in (\ref{SIeq:L4triangle}) by taking its fourth power
\begin{equation}
    f(\vb{x}) = \norm{\vE_i + \sum_{j=1}^q x_j s_j \vb{u}_j}^4
\end{equation}
and bound the resulting quartic polynomial $f(\vb{x})$ over the field DOFs $x_j = \vb{v}_j^\dagger \vP$ using SOS programming. As $q$ increases, the SOS bound on $f(\vb{x})$ converges to a fixed value which is the SERS bound for $f(\vP)$. Larger $q$ leads to larger SOS programs that are more expensive computationally; to illustrate this, we compute bounds to the $f(\vb{x})$ with just the conservation of resistive power imposed, i.e., the SOS program is of the form
\begin{subequations}
    \begin{align}
        & \min_{\substack{t \in \mathbb{R} \\ m(\vx) \text{ is quadratic} } } \quad t \\
        \text{s.t.} \quad & t - ( f(\vb{x}) + m(\vb{x}) C_A(\vb{x}) ) \text{ is SOS}. \label{eq:SOSCstrt}
    \end{align}
    \label{SIeq:AsymSOS}
\end{subequations}
Table \ref{tab:runtime} documents the time taken to solve (\ref{SIeq:AsymSOS}) as a function of $q$ along with the resulting SOS bound of $f(\vb{x})$ for an $E_z$ polarized suspended substrate problem with $L=1.9\lambda$, $H_D = H_R = 0.5\lambda$, $d=0.2\lambda$. The time needed to solve a SOS program with 2 constraints as given by (7) in the main text is generally slower by a factor close to unity. To fully solve (7) in the main text, the outer optimization of the constraint interpolation variable $\gamma$ requires around 10 to 20 SOS solves. We note that in this case the bounds have converged for $q \geq 5$. 

\begin{table}[htp]
\caption{Solve time of the SOS (\ref{SIeq:AsymSOS}) and resulting bound on $f(\vb{x})$ as a function of the mode cutoff $q$. Results computed using 8 cores on a single Intel Xeon Platinum 8260 processor with 16G of RAM. Since each field DOF $x_j$ is a complex number, it is represented by 2 real polynomial variables in the SOS program. \label{tab:runtime}}
\begin{center}
\begin{tabular}{||c c c c||}
\hline
$q$ & \# of variables in SOS & Bound of first term in (\ref{SIeq:L4triangle}) (vacuum FOM) & SOS solve time (s) \\
\hline \hline
1 & 2 & 5.94 & 0.34 \\
3 & 6 & 8.2 & 0.37 \\
5 & 10 & $4.1\times 10^4$ & 3.18 \\
7 & 14 & $4.1\times 10^4$ & 20.3 \\
9 & 18 & $4.1\times 10^4$ & 97.4 \\
11 & 22 & $4.1\times 10^4$ & 367 \\
13 & 26 & $4.1\times 10^4$ & 1143
\end{tabular}
\end{center}
\end{table}